\documentclass[aps,prl,twocolumn,amsmath,amssymb,superscriptaddress]{revtex4-1}
\usepackage{amsmath}
\usepackage[pdftex]{graphicx}
\usepackage{epstopdf}
\usepackage{enumerate}
\usepackage{fancyhdr}
\usepackage{siunitx}
\usepackage{float}
\usepackage{color}
\usepackage{verbatim}
\usepackage{gensymb}
\usepackage{braket}
\usepackage{upgreek}
\usepackage{xspace}
\usepackage{color,soul}
\usepackage{type1cm}
\usepackage{lettrine}
\usepackage{xr}
\newcommand{\umtext}{$\upmu$m\xspace}
\newcommand{\ugtext}{$\upmu$g\xspace}
\newcommand{\ustext}{$\upmu$s\xspace}
\newcommand{\ultext}{$\upmu$L\xspace}
\newcommand{\thirteenC}{$^{13}$C\xspace}
\newcommand{\oneH}{$^{1}$H\xspace}
\newcommand{\mgperml}{mg mL$^{-1}$\xspace}

\begin{document}

\title{Phase-Encoded Hyperpolarized Nanodiamond for Magnetic Resonance Imaging}
\author{David E. J. Waddington}
\affiliation{ARC Centre of Excellence for Engineered Quantum Systems, School of Physics, University of Sydney, Sydney, NSW 2006, Australia}

\author{Thomas Boele}
\affiliation{ARC Centre of Excellence for Engineered Quantum Systems, School of Physics, University of Sydney, Sydney, NSW 2006, Australia}

\author{Ewa Rej}
\affiliation{ARC Centre of Excellence for Engineered Quantum Systems, School of Physics, University of Sydney, Sydney, NSW 2006, Australia}

\author{Dane R. McCamey}
\affiliation{ARC Centre of Excellence for Exciton Science, School of Physics, University of New South Wales, Sydney, NSW 2052, Australia}

\author{Nicholas J. C. King}
\affiliation{The Discipline of Pathology, School of Medical Sciences, Bosch Institute, Sydney Medical School, University of Sydney, Sydney, New South Wales 2006, Australia}

\author{Torsten Gaebel}
\affiliation{ARC Centre of Excellence for Engineered Quantum Systems, School of Physics, University of Sydney, Sydney, NSW 2006, Australia}

\author{David J. Reilly*}
\affiliation{ARC Centre of Excellence for Engineered Quantum Systems, School of Physics, University of Sydney, Sydney, NSW 2006, Australia}
\affiliation{Microsoft Corporation, Station Q Sydney, University of Sydney, Sydney, NSW 2006, Australia}

\begin{abstract}
Surface-functionalized nanomaterials can act as theranostic agents that detect disease and track biological processes using hyperpolarized magnetic resonance imaging (MRI). Candidate materials are sparse however, requiring spinful nuclei with long spin-lattice relaxation ($T_1$) and spin-dephasing times ($T_2$), together with a reservoir of electrons to impart hyperpolarization. Here, we demonstrate the versatility of the nanodiamond material system for hyperpolarized \thirteenC MRI, making use of its intrinsic paramagnetic defect centers, hours-long nuclear $T_1$ times, and $T_2$ times suitable for spatially resolving millimeter-scale structures. Combining these properties, we enable a new imaging modality that exploits the phase-contrast between spins encoded with a hyperpolarization that is aligned, or anti-aligned with the external magnetic field. The use of phase-encoded hyperpolarization allows nanodiamonds to be tagged and distinguished in an MRI based on their spin-orientation alone, and could permit the action of specific bio-functionalized complexes to be directly compared and imaged.
\end{abstract}
\maketitle

\lettrine[lines=3]{D}{}iamond is a remarkable material with outstanding thermal, mechanical, optical, and electrical properties that have given rise to a diverse spectrum of technologies since it was found to be readily synthesizable in the 1950s. In the context of nanomedicine, synthetic nanodiamond can act as a bright, fluorescent marker \cite{Fu2007} of use in sub-cellular tracking of biological processes \cite{Kucsko2013}, \emph{in vivo} sequestering of migratory cell populations \cite{Getts2014} or as a non-toxic substrate for the delivery of chemotherapeutic payloads \cite{Weissleder2014,Almeida2014,Li2014,Boles2016}. Further, the extremely weak nuclear magnetism in diamond makes the material interesting for hosting quantum information \cite{Maurer2012} and for constructing quantum sensors that exploit its long-lived nuclear \cite{Waldherr2012} or electron \cite{Grinolds2013} spin-states.

The long lifetime of these spin-states has recently motivated the development of new nanodiamond-based theranostic agents that are detectable with magnetic resonance imaging (MRI) by acquiring signal from their $^{13}$C nuclei \cite{Rej2015a,Parker2017}. In natural abundance carbon however, spin-1/2 $^{13}$C nuclei comprise only 1.1\% of the diamond lattice, rendering their signal undetectable using conventional MRI. The presence of paramagnetic centers in diamond offers a means of boosting the weak signal from $^{13}$C using hyperpolarization to transfer the much larger thermal polarization of the centers to the $^{13}$C nuclei, by driving spin transitions optically \cite{King2015a}, or with microwaves at cryogenic temperatures \cite{Rej2015a,Dutta2014,Casabianca2011,Bretschneider2016}. Conversely, the presence of paramagnetic centers also shortens the nuclear spin relaxation time ($T_1$), limiting the timescale over which hyperpolarized nuclei can be put to use \cite{Cassidy2013,Atkins2013}. Beyond simply detecting the boosted signal from $^{13}$C, constructing spatially-resolved images requires that the nuclear dephasing time ($T_2$) is sufficiently-long, relative to the timescale set by the strength of practical magnetic field gradients. For this reason solid-state compounds with short $T_2$ times have always presented a major challenge for imaging.\\

\begin{figure*}
	\centering
	\includegraphics[width = 180 mm]{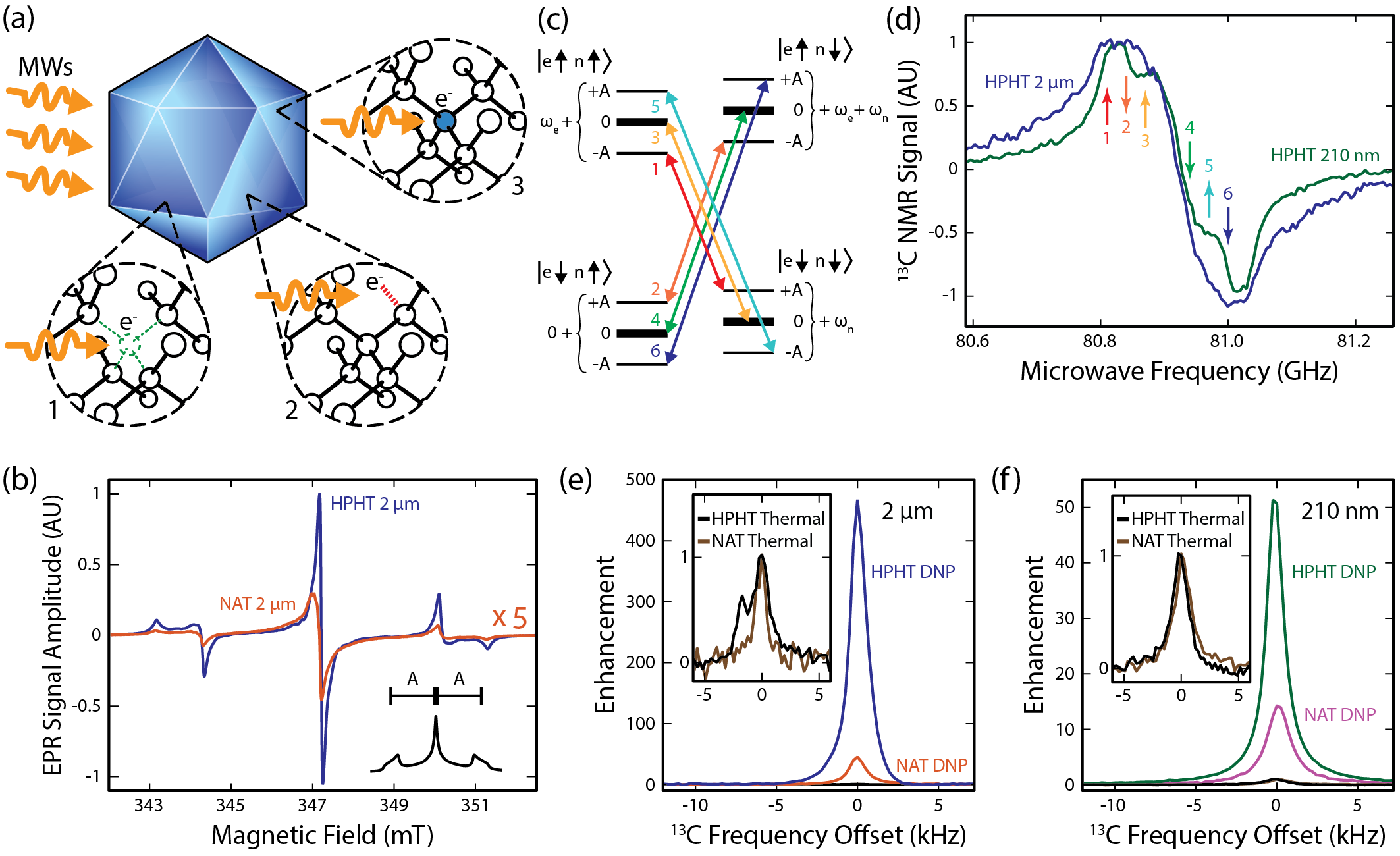}
	\caption{\label{fig1} \textbf{Hyperpolarization of Nanodiamond.} 
		\textbf{(a)} Nanodiamonds (blue) are host to spin-1/2 defects such as charged vacancy sites (1), dangling bonds (2) and substitutional nitrogen atoms (3). Resonant microwaves can be used to drive spin polarization from these defects to $^{13}$C nuclei in the lattice.
		\textbf{(b)} X-band EPR spectra of 2~\umtext HPHT (blue) and 2~\umtext NAT (orange) diamonds. The HPHT and NAT nanodiamonds have substitutional nitrogen impurities present at 95 ppm and 4 ppm respectively. Other spin-1/2 defects are present at concentrations of 115 ppm in the HPHT nanodiamonds and 28 ppm in the NAT nanodiamonds. A representative EPR absorption signal is also shown (inset, bottom right). Hyperfine splitting (A) of the EPR line results from nitrogen impurities.
		\textbf{(c)} Energy level schematic of Zeeman split electrons (e) and nuclei (n), showing possible flip-flip and flip-flop transitions in diamond. Transitions are numbered in order of increasing energy (1-6).
		\textbf{(d)} Hyperpolarization spectra of \thirteenC as a function of microwave frequency for HPHT 210~nm (green) and HPHT 2~\umtext (blue) diamond particles. Signals are normalized to unity. Transitions shown in (c) are indicated by arrows (1 - 80.81 GHz; 2 - 80.84 GHz; 3 - 80.87 GHz; 4 - 80.94 GHz; 5 - 80.97 GHz; 6 - 81.0 GHz).
		\textbf{(e)} \thirteenC NMR spectra of HPHT (blue) and NAT (orange) 2~\umtext diamonds after 20 minutes of hyperpolarization, normalized to the NMR signal from thermal polarization (HPHT - black; NAT - brown).
		\textbf{(f)} \thirteenC NMR spectra of HPHT (green) and NAT (purple) 210~nm nanodiamonds after 20 minutes of hyperpolarization, normalized to the NMR signal from thermal polarization (HPHT - black; NAT - brown).
	}
\end{figure*}

Here we address these challenges by demonstrating the use of nanodiamond for hyperpolarized-$^{13}$C MRI. Via electron paramagnetic resonance (EPR) measurements, we first show that synthetic nanodiamond contains a concentration of impurities and dangling-bonds that is well suited to hyperpolarization (see Fig. \ref{fig1}a), without significantly degrading the long nuclear $T_1$ time. In a major advance over previous work \cite{Rej2015a}, we highlight how hyperpolarization via paramagnetic centers in nanodiamond opens up new avenues for preparing spin-populations with unique MRI contrast signatures. Reasoning that spatially-resolved imaging requires suitably long spin coherence, data characterizing the dephasing times for $^{13}$C nuclei is presented, demonstrating the use of dynamical decoupling pulse sequences that preserve coherence and offer a means of improving image resolution. Testing the viability of hyperpolarized nanodiamond for MRI, we first acquire image data as a function of time, nanoparticle concentration, and size, before deploying the technique in a setting relevant to preclinical applications, acquiring a co-registered $^{13}$C - $^{1}$H image in a mouse \textit{post mortem}.

Taken in combination, the properties of the nanodiamond spin-system enable a new, phase-encoded imaging modality not possible using hyperpolarized liquid compounds with short $T_1$ relaxation times \cite{Cunningham2016}. Extending the repertoire of nanodiamond for MRI \cite{Rej2017,Waddington2017,Rammohan2016}, we show how this method provides an \emph{in situ} control, such that nominally identical nanoparticles can be tagged and distinguished in an MRI by the direction of their nuclear hyperpolarization alone. Finally, we speculate on the use of phase-encoded hyperpolarized MRI for addressing common challenges faced with targeted nanoparticle theranostics. \\

\noindent{\bf{Intrinsic Defects as a Source of Polarized Spins}}
Turning to the details of our experiments, we begin by comparing synthetic nanodiamond produced via the high-pressure high-temperature method (HPHT), to natural nanodiamond (NAT) samples, highlighting the suitability of each for hyperpolarized MRI. Both kinds of nanodiamond exhibit a series of intrinsic electronic defects, immediately apparent in their X-band EPR spectra, shown in Fig. \ref{fig1}b. The generic features in the spectra can be attributed to a convolution of a hyperfine-split P1 center associated with substitutional nitrogen atoms \cite{Barklie1981}, a narrow single-line component from vacancy sites, and a broad single-line that stems from dangling carbon bonds \cite{Loubser1978, Zaitsev2001}. We find that P1 center concentration in the HPHT diamonds (95 ppm) is more than an order of magnitude higher than the P1 center concentration in the NAT nanodiamonds (4 ppm) [see Supp. Fig. \ref{figsupp_epr} and Supp. Table \ref{defect_conc} for discussion of particle size dependent effects]. 

In the presence of an external magnetic field ($B =$~2.88~T), these paramagnetic defects provide a platform for nuclear hyperpolarization by microwave driving the manifold of hyperfine transitions shown in Fig. \ref{fig1}c. Although room temperature hyperpolarization is possible \cite{Rej2015a}, the most significant enhancements occur when microwaves are applied at cryogenic temperatures (4 K or below), transferring the large Boltzmann polarization of the electronic defects to closely coupled \thirteenC nuclei. For spins in nanodiamond, hyperpolarization occurs via a mixture of the solid-effect and cross-effect and leads to a polarized nuclear population of order 8\% at saturation [see Supp. Note \ref{supp:hp_mech} for discussion of polarization mechanism]. Detuning the microwave frequency below or above the central EPR line drives hyperfine transitions that build a positive or negative enhancement in the \thirteenC signal (see Fig. \ref{fig1}c). The sign of the signal indicates nuclear hyperpolarization that is aligned or anti-aligned with the external magnetic field. Hyperpolarizing HPHT and NAT nanodiamond yields the \thirteenC signal enhancements shown in Fig. \ref{fig1}d and \ref{fig1}e, for a mean particle diameter $d$ = 2~\umtext and $d$ = 210~nm respectively. We draw attention to the variation in signal enhancement with particle size and when using nanodiamonds with different concentrations of paramagnetic defects (HPHT versus NAT). \\

\noindent{\bf {Spin Relaxation and Spin Dephasing}}\\
The concentration and configuration of paramagnetic centers in nanodiamond carry-over into the characteristic times for (electron-mediated) nuclear spin relaxation, limiting the time for which hyperpolarization can be usefully deployed. Surprisingly, in their readily available synthetic form, HPHT nanodiamonds are well-optimized for hyperpolarization applications, retaining up to 11 \% of their polarization two hours after being transferred to an imaging platform. Decay rates for the smaller, 210~nm particles are faster but still sufficient for imaging applications, with 10 \% of hyperpolarization remaining after a period of 20 minutes [see Supp. Fig. \ref{figsupp_decay} for data].

\begin{figure}
	\centering
	\includegraphics[width = 88 mm]{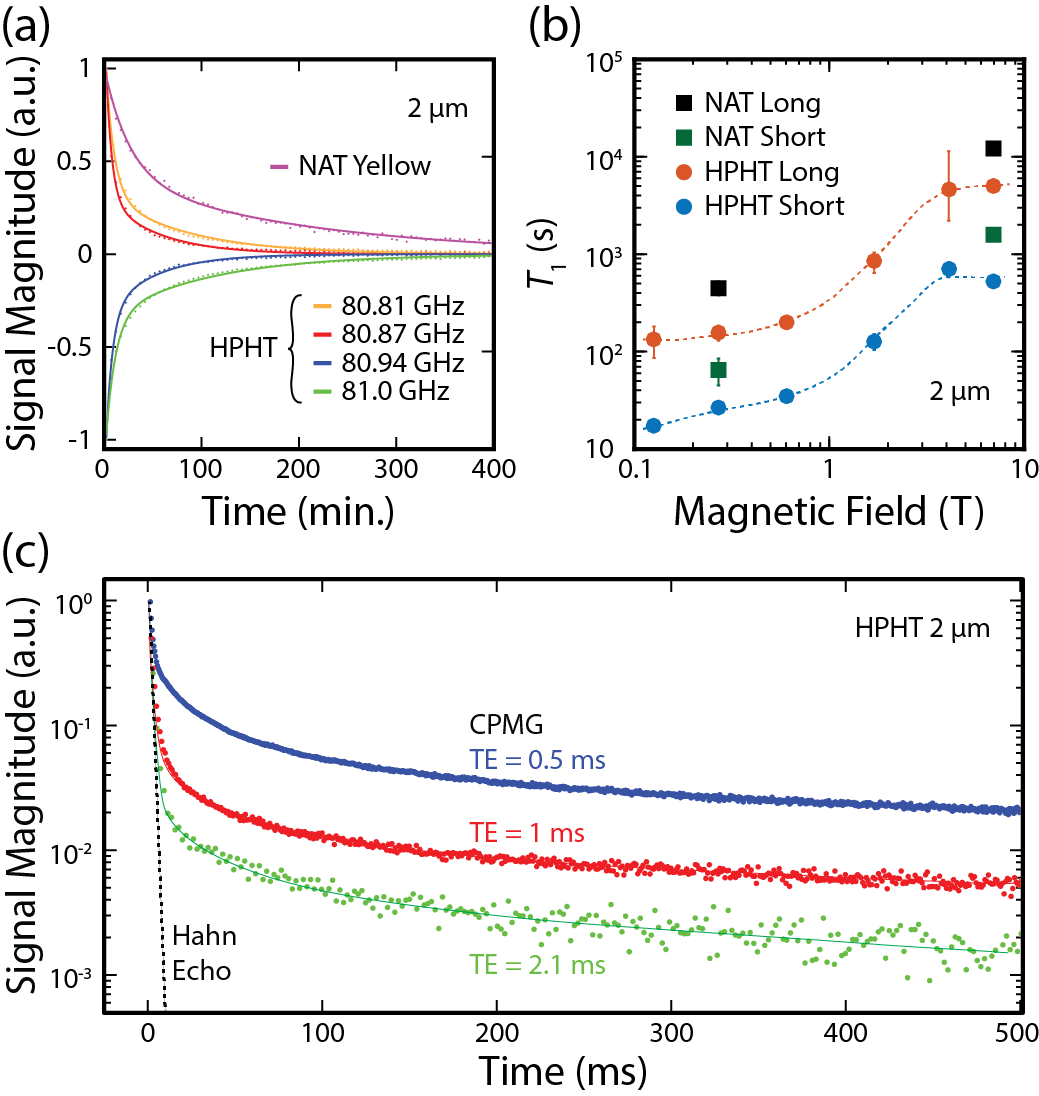}
	\caption{\label{fig2} \textbf{Decay of Hyperpolarization.}
		\textbf{(a)} Signal decay of hyperpolarized 2~\umtext diamond particles at $T = 293$ K and $B_0 = 7$~T. Decay curves are shown for NAT 2~\umtext diamonds hyperpolarized at 80.87 GHz (purple) and HPHT 2~\umtext diamonds hyperpolarized at various frequencies (red - 80.81 , yellow - 80.87 GHz, green - 80.94 GHz and blue - 81.0 GHz). Solid lines are double exponential fits to the data (see Supp. Table \ref{supptab_2umdecay} for fit parameters). 
		\textbf{(b)} $T_1$ relaxation time versus magnetic field for 2~\umtext diamond particles. Circular markers show the HPHT long (red), NAT long (black), HPHT short (blue) and NAT short (green) components of a double exponential fit to the hyperpolarized decay curve (raw data shown in Supp. Fig. \ref{figsupp_decay}). Dashed lines are a guide to the eye.
		\textbf{(c)} Signal magnitude of echoes in a hyperpolarized 2~\umtext diamond sample. CPMG traces are shown with various echo spacings (TE = 0.5~ms - blue; TE = 1~ms - red; TE = 2.1~ms - green). The fit to a Hahn echo experiment performed with thermally polarized 2~\umtext diamond is also shown (black dotted line; data shown in Supp Fig. \ref{figsupp_fse}).}
\end{figure}

Comparing the relaxation data to the hyperpolarization spectra reveals that driving at select microwave frequencies leads to polarization that relaxes with a distinct nuclear $T_1$ (colored arrows in Fig. \ref{fig1}d correspond to colored curves in Fig. \ref{fig2}a). This dependence likely arises due to the microwave frequency selecting particular paramagnetic centers to act as hyperpolarization lattice sites for their surrounding nuclei. The same centers then dominate nuclear spin relaxation, such that different types of centers and their concentration will lead to different relaxation rates.

\begin{figure*}
	\centering
	\includegraphics[width = 180 mm]{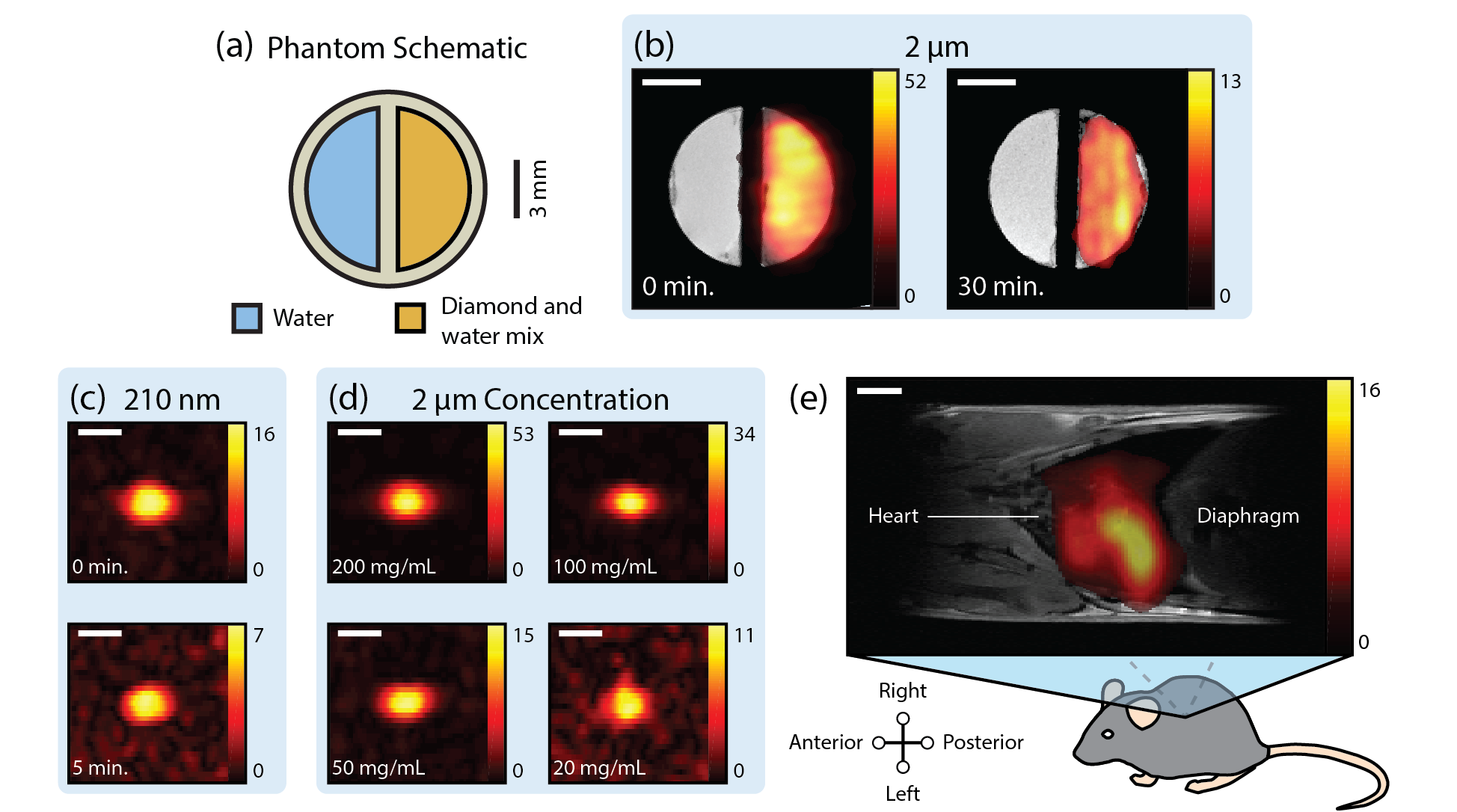}
	\caption{\label{fig3} \textbf{Hyperpolarized Nanodiamond Imaging.} 
		\textbf{(a)} Phantom schematic. One side of the ``half-moon'' phantom was filled with a mixture of hyperpolarized diamond and water (orange). The other half of the phantom is filled with water (blue). 
		\textbf{(b)} Hyperpolarized 2~\umtext microdiamonds were imaged after dissolution in a 120~mg/mL water mixture, using the phantom described in \textbf{a} after 0 minutes and 30 minutes in the scanner.
		\textbf{(c)} Hyperpolarized 210~nm nanodiamonds in a teflon tube at 200~mg/mL concentration were imaged after transfer from the polarizer after 0 minutes and 5 minutes in the MRI scanner. 
		\textbf{(d)} Images of hyperpolarized 2~\umtext microdiamonds in a teflon tube immediately after transfer from the polarizer at various concentrations 
		\textbf{(e)} Co-registered \oneH:\thirteenC MRI of a mouse thorax following intrathoracic injection of hyperpolarized 2~\umtext diamond particles.
		The colorscale in all images indicates \thirteenC signal magnitude in units of the noise floor. All scale bars are 3~mm in length.
	}
\end{figure*}

Examining the decay of the hyperpolarized state shows that the data is bi-exponential, containing short and long characteristic relaxation time components. Further, as a function of magnetic field, both components vary by nearly two orders of magnitude as the field is increased from $B$ = 130~mT to 7~T, as shown in Fig. \ref{fig2}b. We attribute this field dependence to the presence of electron and three-spin mediated processes \cite{Terblanche2001b}, characteristic of nitrogen impurities. Rapid decay of hyperpolarization at low-field presents a practical challenge to the implementation of hyperpolarized nanodiamond imaging, requiring the use of magnetized sample shuttles to ferry compounds between hyperpolarization and imaging platforms \cite{Milani2015} [see Supp. Fig. \ref{figsupp_fieldpath} for details].

Moving from detecting hyperpolarized nanodiamond to imaging clinically-relevant concentrations of the material brings new challenges, in particular, since the short $T_2$ dephasing times of solids can limit the spatial resolution that is possible using typical scanner gradient magnetic field strengths. However, we find the \thirteenC nuclear spins in HPHT nanodiamond fortuitously possess $T_2$ times already suitable for demonstrating the technique, with substantial scope for improving spatial resolution using dynamical decoupling sequences \cite{Li2007}. To see this, in Fig. \ref{fig2}c we compare the signal decay from a standard Hahn-echo sequence (dotted-line) to the signal produced by the use of a Carr-Purcell-Meiboom-Gill (CPMG) decoupling sequence, after the sample has been transfered from the polarizer to a 7~T imaging platform. Although the $T_2$ seen with Hahn-echo is already suitable for sub-millimetre imaging (see Supp. Note \ref{supp:resolution} for discussion of resolution limits), we find that spin-coherence is preserved for a significantly longer time by decreasing the inter-pulse spacing in the CPMG sequence. This enhancement in coherence indicates that repeated $\pi$-pulses have the effect of refocusing the homonuclear dipolar coupling, significantly narrowing the spectral linewidth \cite{Dong2008} with promise to improve the imaging resolution of hyperpolarized nanodiamond.\\

\noindent{{\bf{MRI with Hyperpolarized Nanodiamond}}}\\
Having established nanodiamond as a viable material for hyperpolarization applications, we now turn to the main result of our work, the demonstration of nanodiamond imaging with hyperpolarized \thirteenC MRI. We restrict our imaging experiments to the HPHT diamonds due to their larger signal enhancement. We begin our study by preparing the ``half-moon'' phantom shown in Fig. \ref{fig3}a, which consists of one chamber filled with water and one with an aqueous mixture of 2~\umtext diamond particles that have been hyperpolarized for 2 hours. The magnetic field that the diamond particles were exposed to was maximized at all times during the transfer and dissolution process to retain hyperpolarization [see Methods]. Once transfer of the phantom to the MRI system is complete, \thirteenC imaging is performed immediately with a centrically-ordered, ultrafast gradient echo (GRE) sequence operating at $B$ = 7~T. Sequences such as GRE allow for very short echo times with small tip angles, making them well suited to imaging hyperpolarized dipolar systems such as diamond, where $T_2$ is short and polarization is nonrenewable between phase encoding steps.

Immediately following the acquisition of the $^{13}$C signal we perform conventional \oneH MRI in order to generate a co-registered water-nanodiamond image, shown in Fig. \ref{fig3}b. The \oneH component (gray-scale) of the co-registered image clearly shows the location of water and structure of the phantom, with the \thirteenC component (red-orange) indicating the location of the hyperpolarized nanodiamond solution. We repeat the experiment, now with a delay of 30 minutes between transfer of the phantom to the MRI system and subsequent imaging. Although the image taken after 30 min shows a significant reduction in \thirteenC signal (down to $\sim$ 25 \%, of the image that was acquired with no delay), we note that it still contains sufficient contrast to be able to identify the presence of nanodiamond. Switching 
to the smaller 210~nm nanodiamonds yields images with a reduced signal-to-noise (SNR) (comparable to 30\% of the 2~\umtext solution), with the images decaying more rapidly in keeping with the size dependence of $T_1$ [see Fig. \ref{fig3}c and Supp. Table \ref{t1table}].

Evaluating the sensitivity of \thirteenC MRI in our current setup, we acquire images for aqueous solutions of hyperpolarized nanodiamonds in a range of concentrations from 20 - 200~\mgperml, as shown in Fig. \ref{fig3}d. For the larger 2~\umtext particles, concentrations as low as 20~\mgperml continue to yield an acceptable SNR $\sim$ 11, corresponding to 57 \ugtext of diamond per pixel. This particle mass sensitivity is similar to other particle imaging techniques based on hyperpolarization \cite{Cassidy2013,Waddington2017}. \\

\noindent{\bf{Animal Imaging}}\\
Aware that MRI phantoms can artificially enhance image quality, we demonstrate hyperpolarized nanodiamond imaging in a setting that is more realistic to the clinic by acquiring a co-registered \thirteenC:\oneH MRI of a \emph{post mortem} three-week old laboratory mouse (\emph{mus musculus}). Using 2~\umtext HPHT nanodiamond, the sample was hyperpolarized for 2 hours before rapid dispersion into a syringe incorporating a 450 mT Halbach array to retain hyperpolarization. The mouse was positioned on a stage within an open-access permanent magnet and the hyperpolarized bolus then injected into the thoracic cavity. Transfer of the mouse to the 7~T MRI system again made use of a custom Halbach array. The total time between removal of the sample from the hyperpolarizer and insertion of the mouse in the MRI system was $\sim$ 60 seconds.

\begin{figure*}
	\centering
	\includegraphics[width = 180 mm]{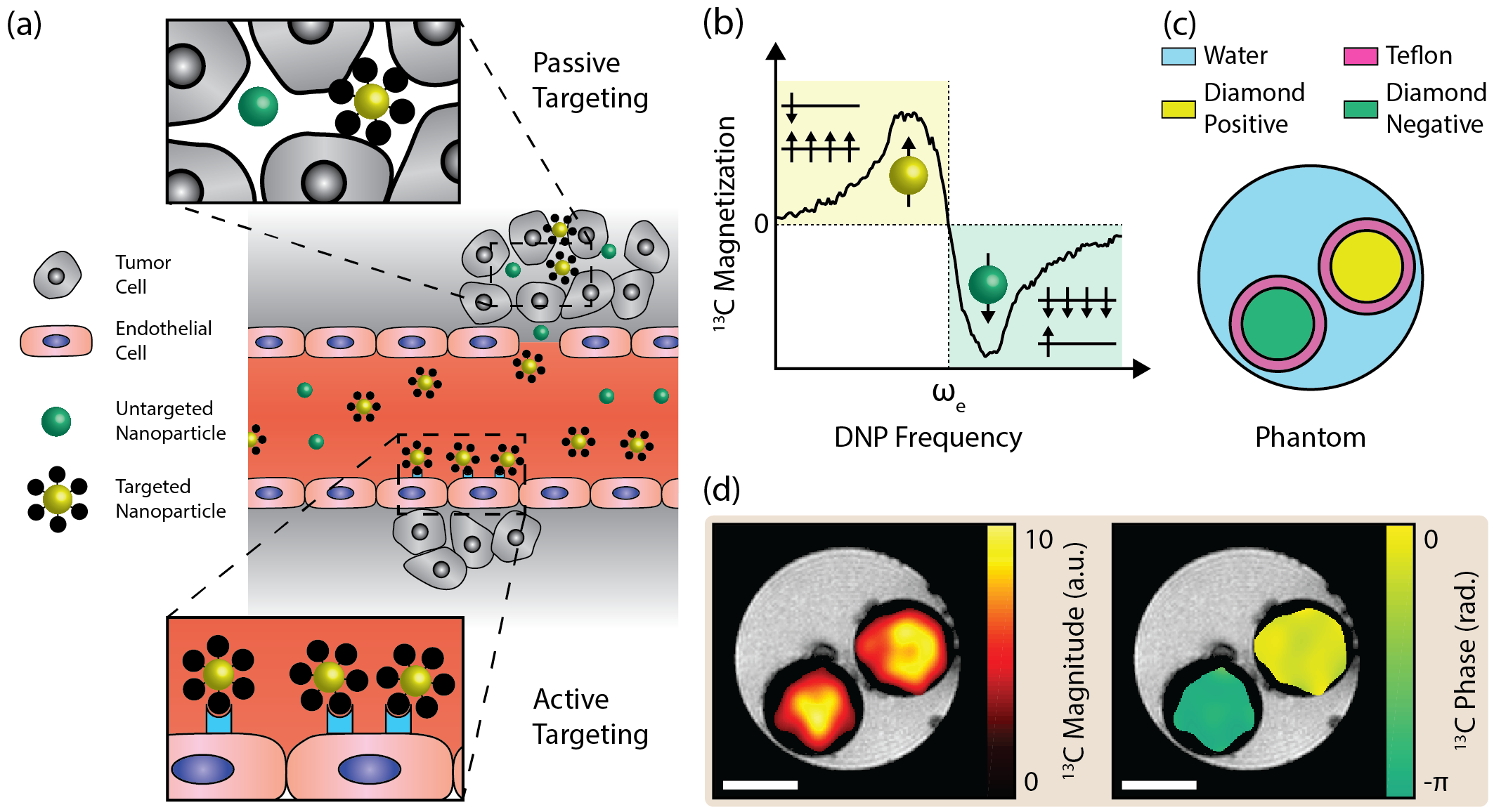}
	\caption{\label{fig4} \textbf{Distinguishing Identical Nanoparticles with Phase-Encoded Hyperpolarization.}
	    \textbf{(a)} Schematic representation of functionalized nanoparticles targeting tumors both actively and passively. Active targeting in the vasculature relies on specific interactions between ligands on the nanoparticle surface and receptors expressed on the surface of, for instance, endothelial cells adjacent to tumors. Passive targeting is nonspecific to nanoparticle functionalization and can occur via the enhanced permeability and retention effect in regions of leaky vasculature at tumor sites. Potentially, phase-encoded hyperpolarization can distinguish between these two regimes.
		\textbf{(b)} Direction of hyperpolarized $^{13}$C magnetization versus the pump frequency.
		\textbf{(c)} Phantom schematic showing two 2~\umtext diamond samples (yellow and green) in a tube of water (blue). Teflon walls of the tubes containing diamond are also shown (pink). The sample shown in yellow was hyperpolarized at 80.87 GHz ($f = \omega_\mathrm{e} - \omega_\mathrm{n}$) for 20 minutes before transfer to the imager. The diamond sample shown in green was subsequently hyperpolarized at 80.94 GHz ($f = \omega_\mathrm{e} + \omega_\mathrm{n}$) for 4 minutes before transfer to the hyperpolarizer.
		\textbf{(d)} Imaging of the phantom described in \textbf{b}. Overlaid \thirteenC magnitude (left) and phase (right) colormaps are shown with masks applied to regions with less than 8 times the root mean square value of the background signal. Small black spots in gray-scale \oneH magnitude image are bubbles adhered to surfaces in the phantom. Scale bar is 3~mm in length.
  }
\end{figure*}

Co-registered \oneH:\thirteenC images of nanodiamond in the mouse are shown in Fig. \ref{fig3}e. The \oneH image reveals anatomical features within the chest at sub-millimeter resolution. The \thirteenC diamond image then clearly shows that the nanodiamond is confined to a region around the lungs. We would expect this localization to occur after injection into the thoracic cavity as the 2~\umtext particles are too large to cross membranes into other regions of the mouse \cite{Cassidy2013,Whiting2016}.\\

\noindent{\bf{Phase-Encoded Hyperpolarization}}\\
The use of nanoparticles for future theranostic applications hinges on the effectiveness of their functionalized surfaces, which, for instance, can enable targeted up-take of particles by binding to specific cells and localizing at specific disease foci. Accompanying these active targeting mechanisms are passive biological responses to nanoparticles based on their size, morphology, or particle composition \cite{Sykes2014,Kunjachan2014,Colombo2016,Neri2005}. Having demonstrated hyperpolarized nanodiamond MRI, we now look to the future and present a new imaging modality enabled by the material properties of nanodiamond and which aims to directly address the challenge of resolving passive response mechanisms from the active targeted delivery of nanoparticles. 

The goal of distinguishing actively targeted vectors from a passive uptake is illustrated in Fig. \ref{fig4}a, where the yellow particles feature surfaces with receptor molecules in contrast to the bare nanoparticles shown in green. The challenge then is to establish an imaging modality that enables contrast between nominally identical nanoparticles, based on a labeling mechanism that does not require a difference in particle size, composition, morphology, or physical state. 

A means of tagging nanodiamond populations is evident in the data presented in Fig. \ref{fig4}b [reproduced from Fig. \ref{fig1}c], which shows the preparation of hyperpolarized states with equal magnitude, both aligned, or anti-aligned, with the external magnetic field by selecting the frequency of microwave driving. In each case the populations of \thirteenC spins are prepared predominantly in their ground or excited state, relaxing back to thermal equilibrium on a timescale $T_1$, as indicated in Fig. \ref{fig4}b. 

Imaging contrast between the ground and excited state populations is thus encoded in the opposing phases, rather than amplitudes of the nuclear magnetization. Demonstrating this modality, we hyperpolarize two identical samples of 2~\umtext nanodiamond, the first at frequency $f = \omega_\mathrm{e} - \omega_\mathrm{n}$ and the second at $f = \omega_\mathrm{e} + \omega_\mathrm{n}$, flipping the phase of the magnetization as indicated in the phantom shown in Fig. \ref{fig4}c. Following adiabatic transfer to the imaging platform, we acquire co-registered \oneH:\thirteenC magnitude and \thirteenC phase images, as shown in Fig. \ref{fig4}d. Conventional images constructed from the magnitude of the \thirteenC signal clearly show the location in the phantom of the hyperpolarized nanodiamond samples, but without appreciable contrast between them. Alternatively, significant differential-contrast is apparent in the phase response of the \thirteenC signal, reflecting the phase-labeling of the hyperpolarized states and enabling the nominally identical nanodiamond populations to be readily distinguished.

Simultaneous administration of positively-polarized nanodiamonds, functionalized with active targeting ligands, and negatively-polarized, untargeted diamonds, opens the prospect of tracking and imaging nanodiamonds accompanied by an \emph{in vivo} control. As the targeting ability of functionalized nanoparticles can often disappear when placed in a biological environment \cite{Salvati2013,Kreyling2015}, the ability to non-invasively verify, via MRI phase, that nanoparticle accumulations are due to active targeting appears useful.\\

\noindent{\bf Discussion}\\
Opportunities exist to significantly improve the concentration sensitivity of nanodiamond MRI reported here. At present, transfer of the sample from hyperpolarizer to imaging platform results in a substantial loss of polarization, stemming from the rapid relaxation that occurs in a low-field ($<$ 380 mT) environment. The use of high-field sample-shuttles during transfer can likely increase the signal by a factor of 10 or more [see Supp. Fig. \ref{figsupp_polretain}]. Further, we draw attention to the use of dynamical decoupling pulse sequences in preserving coherence in dilute spin systems such as diamond, as shown in Fig. \ref{fig2}c. Whilst CPMG based imaging sequences have displayed similar sensitivity in our scanner, they offer strong potential for improving spatial resolution and sensitivity when used in conjunction with appropriate gradient-field protocols [see Supp Fig. \ref{figsupp_fse}]. Developing a single-shot sequence based on quadratic echo imaging \cite{Frey2012}, for instance, would allow rapid pulsing and high-resolution imaging via the long echo tails observed in the hyperpolarized nanodiamond system.
 
Perhaps the most impactful opportunity for improving the performance of nanodiamond MRI is in isotopic engineering of the precursor carbon. Enriching the \thirteenC content, from the 1.1\% that occurs in natural abundance, to a few percent will lead to a proportional increase in MRI sensitivity \cite{Balasubramanian2009,Parker2017} with limited impact on relaxation times [see Supp. Note \ref{supp:enrichment}]. A further avenue for improvement is in optimizing the type and concentration of paramagnetic centers in an effort to lengthen spin dephasing times \cite{Knowles2014} and suppress low-field relaxation \cite{Terblanche2001b,Lee2011}. This will be particularly important in extending the applications for smaller nanodiamonds. The use of photo-excited radicals, such as those that persist at low temperatures but recombine on warming the sample \cite{Capozzi2017}, may further provide a means of achieving long-lived hyperpolarization.

In conclusion, we have demonstrated \thirteenC magnetic resonance imaging using nanodiamonds, including the introduction of a new imaging modality based on phase-encoded hyperpolarization, exploiting the long relaxation times inherent to the diamond material system. Future improvements in materials, hardware, and sequence design will likely see the nanodiamond platform emerge as a valuable theranostic tool for multimodal imaging, combining MRI and optical fluorescence to span sub-cellular to whole-body scales.

\section*{Methods}
\textbf{Diamond Particles:}
Nanodiamonds used in this work were purchased from Microdiamant (Switzerland). Specific nanodiamond types were monocrystalline, synthetic HPHT and NAT particles. Both types were used in sizes of 210~nm (0-500~nm, median diameter 210~nm) and 2~\umtext (1.5-2.5~\umtext, median diameter 2~\umtext). The 2~\umtext HPHT diamonds used here are well suited to injection as they have a zeta potential of -38$\pm$7~mV in water. This zeta potential shows that they are aggregation-resistant, a characteristic important to benefiting fully from the surface of nanodiamond in biological applications \cite{Mochalin2012}. The 2~\umtext particles display sedimentation on the timescale of hours, as is to be expected for diamond particles larger than 500~nm, which is the point at which gravitational forces overcome Brownian motion. 210~nm HPHT nanodiamonds have a near identical zeta potential of -39 $\pm$ 8 mV with limited sedimentation observed over a period of weeks. For details of dynamic light scattering measurements and calculation of particle stability, see Supp. Note \ref{supp:sedimentation}.

\textbf{EPR Characterization:}
EPR spectra were measured using a Bruker EMX EPR Spectrometer operating at 9.735 GHz and room temperature. Resulting spectra were fit using Easyspin to a three spin component model \cite{Rej2015a,Stoll2006}. Defect concentrations were calculated relative to the signal from an irradiated quartz EPR standard \cite{Eaton2010}.

\textbf{Hyperpolarization of diamond:}
Hyperpolarization is performed in a 2.88~T superconducting NMR magnet with a homebuilt DNP probe \cite{CassidyThesis}. 80-82 GHz microwaves are generated via a frequency multiplier (Virginia Diodes) and 2 W power amplifier (Quinstar) before being directed to the sample via a slotted waveguide antenna. A helium flow cryostat (Janis) was used to cool the sample to 4.5 K. NMR measurements in the polarizer were taken with a custom saddle coil tuned to the $^{13}$C Larmor frequency of 30.912 MHz. All hyperpolarization events begin with a $^{13}$C NMR saturation sequence ($64 \times \pi/2$ pulses) to zero the initial polarization. Individual points in the frequency sweeps of Fig. \ref{fig1}d correspond to unique hyperpolarization events, showing the $^{13}$C NMR signal from a $\pi/2$ pulse after 60~s of microwave saturation at a constant frequency. NMR enhancement plots in Fig. \ref{fig1}e and 1f show the \thirteenC signal magnitude after 20 minutes of microwave saturation compared with the signal seen after an identical period of thermal polarization. Absolute polarization values are calculated from the thermally polarized \thirteenC signal at equilibrium and the expected Boltzmann polarization \cite{Rej2015a}. Diamond samples were hyperpolarized in teflon tubes, chosen for their transparency to microwaves and robustness to thermal cycling.

\textbf{Transfer and dissolution:}
All post-transfer NMR signals were acquired in the same 7~T superconducting NMR magnet used for imaging. Adiabatic sample transfer between the DNP probe and the MRI scanner occurs via a series of ``magnetic shields'' based on permanent magnets and Halbach arrays (see Supp. Fig. \ref{figsupp_fieldpath} and Supp. Note \ref{supp:adiabatic} for further detail). Halbach arrays are especially useful for transferring samples between superconducting magnets as they experience no net translational force in external magnetic fields. 

Depolarization at 7~T was measured via small tip angle for 2~\umtext and 210~nm samples hyperpolarized for 2 hours and 1 hour respectively. Data were divided by $\left( \cos {\alpha} \right)^{ n-1 }$, where $\alpha$ is the tip angle and $n$ the pulse number, to account for RF-induced signal decay.

$T_1$ versus magnetic field strength was measured in the stray field of the imaging magnet with samples hyperpolarized for 20 minutes. Small tip angle measurements of the \thirteenC signal magnitude were taken at 7~T before and after shuttling the sample to a lower field region for repeated periods of 30 s.

The phantom in Fig. \ref{fig3}a was prepared in an approximately 200 mT permanent magnet by rapid thawing of a \thirteenC hyperpolarized 2:1 mixture of nanodiamond and water. The hyperpolarized mixture was then mixed with additional water to give a 120~\mgperml concentration. Transfer to the MRI system occurred via a 110 mT Halbach array.

\textbf{MRI Experiments:}
All MRI was performed in a 7~T widebore, microimaging system with \oneH and \thirteenC Larmor frequencies of 299.97 MHz and 75.381 MHz respectively. The microimaging gradient set produces gradient fields up to 250 mT m$^{-1}$. Phantoms were imaged using a 10~mm, dual resonance \oneH:\thirteenC NMR probe. \oneH phantom images were acquired with a GRE sequence with 60~\umtext $\times$ 60~\umtext pixel size and 6~mm slice thickness. \thirteenC slice thickness was restricted to 20~mm by the active region of the detection coil. Concentration phantoms contained 140 \ultext of diamond and water mixture.

\thirteenC GRE image data in Fig. \ref{fig3}b-d and \ref{fig4}d were acquired with a 64 $\times$ 32 matrix and pixel resolution of 0.7~mm $\times$ 0.6~mm. The refocusing time (TE) of 1.22~ms was minimized by ramping gradients to full strength and using short, 60 \ustext, excitation pulses (see Supp. Fig. \ref{figsupp2} for complete timing parameters). Centrically-ordered phase encoding with a constant flip angle of 20\degree\xspace was used to increase SNR by 1.96 times at the expense of limited blurring in the phase encode direction \cite{Zhao1996}.

All images displaying \thirteenC data only were interpolated to 128 $\times$ 128 resolution by zero-filling and Gaussian filtering in $k$-space. Co-registered \thirteenC images were interpolated to the resolution of the accompanying \oneH image and values smaller than 3 times (Fig. \ref{fig3}b) or 5 times (Fig. \ref{fig3}e, Fig. \ref{fig4}d) the SNR made transparent to reveal the underlying structure in the \oneH image.

\textbf{Animal Imaging:}
A 150~mg sample of 2~\umtext nanodiamonds was hyperpolarized for 2.5 hours in a 1:1 mixture with water with 0.2 mL of hot water added during dissolution. Small tip angle characterization showed that hyperpolarizing in water or ethanol gave the same \thirteenC enhancement seen for dry samples. The resulting 0.5 mL bolus was injected into the thoracic cavity of a three-week old female mouse \emph{post mortem} (11 g, \emph{mus musculus}) with a 21G needle and syringe before transfer to the MRI system. Hyperpolarized nanodiamonds were magnetically shielded with custom magnets when not in the polarizer or imager (further details in Supp. Fig. \ref{figsupp_exvivomagnets}) . Total time for dissolution, injection and transfer to the imager was approximately 60 s.

A \oneH:\thirteenC double resonance microimaging probe with 18~mm bore size was used for imaging the mouse torso. \thirteenC SNR of this probe was measured to be one third of the 10~mm probe used for phantom imaging. A conventional \oneH spin-echo sequence was used for high-contrast anatomical imaging. \oneH images were acquired with 1~mm slice thickness with a 256 $\times$ 256 matrix and pixel resolution of 230~\umtext $\times$ 176~\umtext. The same \thirteenC GRE sequence was used as in Fig. \ref{fig3}b-d but with a gradient echo time of 0.90~ms and pixel size of 2.1~mm $\times$ 1.9~mm. Slice thickness was limited to less than 18~mm by the sensitive region of the probe. 

\textbf{Phase-contrast Imaging:}
To demonstrate phase-contrast imaging, a 2~\umtext HPHT sample was positively polarized with 80.87 GHz microwaves for 20 minutes and transferred to the MRI scanner. A second, nominally identical sample was then negatively polarized with 80.94 GHz microwaves for 4 minutes and transferred to the MRI scanner.


\section*{Additional Information}

\subsection*{Acknowledgements}
The authors thank M. S. Rosen for useful discussions and M. C. Cassidy for constructing the hyperpolarization probe. We are also grateful to R. Marsh and A. Itkin for the supply of custom hardware for these experiments. For EPR measurements, the authors acknowledge use of facilities in the Mark Wainwright Analytical Centre at the University of New South Wales. This work was supported by the Australian Research Council Centre of Excellence Scheme (Grant No. EQuS CE110001013) and ARC DP1094439. We also acknowledge use of the nanoparticle analysis tools provided by the Bosch Institute Molecular Biology Facility at the University of Sydney.



$^*$Correspondence and requests for materials should be addressed to:\\ D.J.R. (email: david.reilly@sydney.edu.au).

\end{document}